# Simulations of dispersion and deposition of coarse particulate matter

**Rodolfo G. Cionco** [1], **Nancy E. Quaranta** [1,2] **y Marta G. Caligaris** [1]
(1) Facultad Regional San Nicolás, UTN, Colón 332, San Nicolás, Buenos Aires, Argentina; (2) Comisión de Investigaciones Científicas de la provincia de Buenos Aires. E-mail: gcionco@frsn.utn.edu.ar

ABSTRACT

In order to study the dispersion and deposition of coarse anthropogenic particulate matter (PMc, aerodynamic diameters> 10 μm), a FORTRAN simulator based on the numerical integrator of Bulirsch and Stoer has been developed. It calculates trajectories of particles of several shapes released into the atmosphere under very general conditions. This first version, fully three-dimensional, models the meteorology under neutral stability conditions. The simulations of such pollutants are also important because the standard software (usually originating in the United States Environmental Protection Agency-EPA-) describe only the behavior of PM10 (diameter less than 10 μm). Bulirsch and Stoer integrator of widespread use in astrophysics, is also very fast and accurate for this type of simulations. We present 2D and 3D trajectories in physical space and discuss the final deposition in function of various parameters. PMc simulations results in the range of 50-100 μm and densities of 5.5 g cm$^{-3}$ emitted from chimneys, indicate that for the purpose of deposition, the emission velocities are not as important as wind speed and the shape of the particles. For densities less than 2 g cm$^{-3}$, the lack of consideration of buoyancy introduces important changes in the distribution of deposited PMc.

**RESUMEN**

Con el objeto de estudiar dispersión y deposición de material particulado grueso antropogénico (PMc, diámetros aerodinámicos > 10 μm), se ha desarrollado un simulador FORTRAN basado en el integrador numérico de Bulirsch y Stoer, que calcula trayectorias de partículas de diversas formas liberadas al ambiente bajo condiciones muy generales. Esta primera versión, totalmente tridimensional, modela la meteorología en circunstancias de estabilidad neutral. La presentación de un simulador de dispersión de este tipo de contaminantes es, además, relevante debido a que los software de uso corriente (generalmente originados en la United States Environmental Protection Agency –USEPA-) describen sólo el comportamiento del $PM_{10}$ (diámetros menores que 10 μm). El integrador Bulirsch y Stoer, de difundido uso en astrofísica, resulta también muy rápido y preciso para este tipo de simulaciones. Se presentan trayectorias 2D y 3D en el espacio físico y se analiza la deposición en función de diversos parámetros. Resultados de simulaciones de PMc en el rango de 50-100 μm y densidades de 5.5 g cm$^{-3}$ emitidos por chimeneas indican que, a efectos de la deposición, las velocidades de emisión del PMc no son tan importantes frente a la velocidad del viento y la forma de las partículas. Para densidades menores que 2 g cm$^{-3}$, la no consideración de la flotabilidad introduce importantes modificaciones en la distribución del PMc depositado.

## INTRODUCCIÓN

El estudio del material particulado atmosférico antropogénico (PM), es uno de los temas de mayor actualidad e importancia de las ciencias medioambientales, particularmente en el estudio de subproductos industriales (Chow y Watson, 2008). PM es el nombre genérico que designa a *partículas* de tamaños variables (desde nanométricas hasta de unos cien micrones de diámetro) y diferente composición, que se liberan a la atmósfera mediante diversos procesos artificiales (industria, transporte, generación de energía, etc.). De estas partículas, aquellas cuyos diámetros *aerodinámicos* son menores que 10 μm (i.e., con velocidades de sedimentación iguales a las de una esfera de 10 μm de diámetro y 1 g cm$^{-3}$ de densidad) son las más nocivas para la salud, ya que permanecen más tiempo en suspensión y pueden llegar con facilidad hasta los alvéolos pulmonares (Asgharian et al., 2006); por lo tanto, se han puesto los mayores esfuerzos en el modelado de la dispersión de esta clase de PM (USEPA, 2009). Sin embargo, el PMc forma parte de los procesos generales de contaminación sobre el medioambiente circundante a las zonas de emisión contribuyendo, por ejemplo, con trastornos del tracto respiratorio superior (Chow y Watson, 2007; Pairon y Roos, 2001). En suma, el estudio de la dispersión y deposición de PMc es imprescindible para una descripción teórica completa del flujo de partículas contaminantes en un determinado sitio. Vesovic et al. (2001), es la referencia más difundida que ha puntualizado la necesidad de un mejor entendimiento y cuantificación del transporte y deposición de este tipo de material. Estos autores estudian PMc de diámetros entre 75 y 106.7 μm, comparando las concentraciones predichas por un clásico modelo de pluma gaussiana (Fugitive Dust Model), con un modelo desarrollado por ellos para resolver las ecuaciones de movimiento de las partículas liberadas en la atmósfera bajo condiciones de estabilidad neutral. Este modelo computacional está basado en un método de diferencias finitas con paso ajustable. El resultado de Vesovic et al. es concluyente: el modelo gaussiano introduce sobreestimación en el pico de concentración y subestimación de la concentración de contaminantes corriente abajo, en forma estadísticamente significativa. En efecto, el PMc (que también podría denominarse PM *pesado*), está dominado por la gravitación y la interacción con la atmósfera mediante un arrastre de Stokes a bajos números de Reynolds (Seinfeld, 1985), por lo tanto, no puede ser tratado como un fluido. Además Vesovic et al., puntualizan que su código de dispersión es bidimensional debido al costo computacional que les supone evaluar las trayectorias en una tercera dimensión. Si bien la justificación para usar un código 2D es razonable debido a que las partículas básicamente van a seguir la dirección de la corriente principal del viento, la aproximación no es válida en situaciones donde los ejes del sistema de referencia utilizado no coincidan con la dirección del viento, ésta presente rotaciones o la intensidad de la turbulencia (siempre presente en todas direcciones) produzca sensibles apartamientos de las partículas respecto a la corriente principal del viento. El trabajo de Vesovic et al. es también limitado desde el punto de vista que no involucra partículas emitidas por chimeneas industriales.
En este trabajo se presenta la implementación de un código FORTRAN para el cálculo preciso y rápido de trayectorias tridimensionales de PMc. En las páginas siguientes se describen las principales características del modelo implementado y el código desarrollado, se discuten comparaciones con el trabajo anteriormente citado y ejemplos concretos de aplicación.

## MODELO MATEMÁTICO

Para determinar las trayectorias del PMc liberado desde chimeneas industriales, en condiciones de estabilidad neutral, deben integrarse las ecuaciones de movimiento bajo la acción de la gravedad terrestre y la interacción con la atmósfera, todo esto, en ausencia de grandes movimientos verticales en la masa de aire circundante. Sea $\vec{r} = \vec{r}(t) = (x, y, z)$, la posición de una partícula en un sistema de referencia fijo en Tierra; salvo indicación en contrario, se dirigirá el eje *X* según la dirección horizontal coincidente con la corriente principal del viento; el eje *Z* coincidente con la vertical del lugar (solidario con el eje longitudinal de la chimenea) y el eje *Y* completa la terna directa. $\vec{v} = \dot{\vec{r}}(t)$ es la velocidad de la partícula; $\vec{u} = (u_x, u_y, u_z)$ la velocidad del viento. Considerando las fuerzas actuantes (gravitación, flotación y arrastre de Stokes), la aplicación de la segunda ley de Newton provee:

$$\frac{d^2\vec{r}}{dt^2} = -\left(1 - \frac{\rho_l}{\rho_p}\right)\vec{g} + \frac{f}{\tau}(\vec{u} - \vec{v}), \quad (1)$$

donde $\rho_l$ es la densidad del fluido; $\rho_p$ es la densidad de la partícula; $\vec{g}$ es la aceleración de la gravedad; además:

$$f = 1 + 0.15\,Re^{0.687} \quad (2)$$

es una función del número de Reynolds (*Re*) y proviene del régimen de Stokes para *Re* bajos (< 200) que son los de interés para este trabajo (Göz et al., 2004; Seinfeld, 1985);

$$\tau = \frac{m_p}{3\pi\mu D_p} \quad (3)$$

es el tiempo de relajación para una partícula en un régimen perfectamente laminar ($m_p$ es la masa de la partícula, $D_p$ es su diámetro y $\mu$ la viscosidad dinámica del fluido); $\tau$ es el tiempo que tarda la partícula en perder su memoria dinámica y responder ante cambios de velocidad en el fluido ideal. El viento se modela, como es usual, considerando un viento medio:

$$\vec{u}_m = \frac{\vec{u}_*}{\kappa} \ln\left(\frac{z+z_0}{z_0}\right) \tag{4}$$

donde $\vec{u}_*$ es la velocidad de cizalla (con componentes *X* e *Y*), $\kappa$ es la constante de von Karmann y $z_0$ la altura de la capa límite (definida a partir de las irregularidades del terreno). También se han incluido fluctuaciones turbulentas a la velocidad ($\vec{u}_t$), representadas en sus tres componentes, como soluciones de la ecuación de Langevin:

$$\vec{u}_t(t,z) = \vec{u}_t(t-dt,z)\ \exp\left(-\frac{dt}{\tau_L}\right) + \vec{\sigma}\ rnd\left(1-\exp\left(-\frac{2dt}{\tau_L}\right)\right)^{1/2} \tag{5}$$

donde, *dt* es el lapso de tiempo en el cual se evalúa la turbulencia, *rnd* es un número aleatorio que se genera con distribución gaussiana en (-1,1), $\tau_L$ es la escala de tiempo lagrangiana y $\vec{\sigma}$ es la dispersión de velocidades, ambas definidas por las siguientes expresiones

$$\tau_L = \frac{0.4\,z}{|\vec{u}_*|} \tag{6}$$

$$\vec{\sigma} = (2.4|\vec{u}_*|, 2.4|\vec{u}_*|, 1.25|\vec{u}_*|) \tag{7}$$

## EL CÓDIGO

La Ec. (1) se integra mediante el método de Bulirsch y Stoer (Press et. al 1992). Para ello se realizó un programa FORTRAN que utiliza la rutina BSSTEP de Numerical Recipes (Press et al., 1992). El método usa extrapolación de Richardson para aproximar la función a integrar mediante una subdivisión automática y arbitrariamente pequeña del paso de integración máximo inicial ($h_1$), de tal forma que se satisfaga una tolerancia (EPS) prefijada. Este proceso garantiza gran precisión con un costo computacional mínimo. El programa comienza leyendo un archivo de condiciones iniciales de posición y velocidad para una partícula; el usuario introduce un número semilla para la generación de la secuencia de pseudoaleatorios (Ec. 5), los cuales se generan mediante un procedimiento de L´Ecuyer (Press et al., 1992). La Ec. (1) se descompone en un sistema de seis ecuaciones acopladas; en cada paso de integración se calculan las componentes deterministas de las fuerzas intervinientes y las estocásticas (turbulencia), evaluándose las condiciones aerodinámicas y meteorológicas respectivas. Las salidas dan posición y velocidad de las partículas cada cierto intervalo de tiempo fijado por el usuario, el tiempo de vuelo y las coordenadas y velocidades en el punto de impacto. Una vez que la partícula llega hasta la altura mínima de integración (ZMIN, generalmente = $z_0$) se imprimen las salidas y el programa reinicia, a partir de las condiciones iniciales, generando para una partícula idéntica un ambiente turbulento distinto. De esta forma el código funciona mediante un procedimiento Monte Carlo que permite seguir hasta cierta altura sobre el suelo un número arbitrario de partículas. La validación del programa se hizo mediante diversas comparaciones con problemas de disparo de proyectiles totalmente integrables (viento con velocidad constante).

## APLICACIONES y RESULTADOS

Una primera aplicación ha sido cotejar con resultados de Vesovic. et al (2001) sobre experimentos de dispersión de esferas de vidrio liberadas desde alturas máximas de 15 m. A modo de ejemplo, se presentan comparaciones con dos casos emblemáticos del trabajo. En primer lugar se simuló la dispersión de esferas de 106.7 µm de diámetro liberadas desde 15 m donde la velocidad del viento fue de 7.31 m $s^{-1}$, (cálculo puramente 2D). Estos autores no dan detalles del resto de las condiciones iniciales ni demás parámetros de integración ni del número de partículas utilizadas. A continuación se presenta la simulación de $10^3$ partículas que se siguen hasta una altura ZMIN = 1.5 m; de acuerdo a esto (Ec. 6), el paso de integración $h_1$ utilizado debería ser del orden de 0.4 s (valor adoptado). La Fig. 1 (a) muestra las trayectorias (2D) de algunas partículas simuladas. La Fig. 1 (b), muestra el histograma de deposición que debe compararse cualitativamente con la Fig. 1, de Vesovic et al. (2001), resultando muy buen acuerdo entre ambas. Comparaciones cuantitativas se realizaron con experimentos de dispersión de partículas de 100 y 75 µm, para las cuales los autores dan estimaciones de pico central de concentración, mediana y otros parámetros de dispersión. Los resultados (no mostrados por brevedad) son también altamente satisfactorios encontrándose las

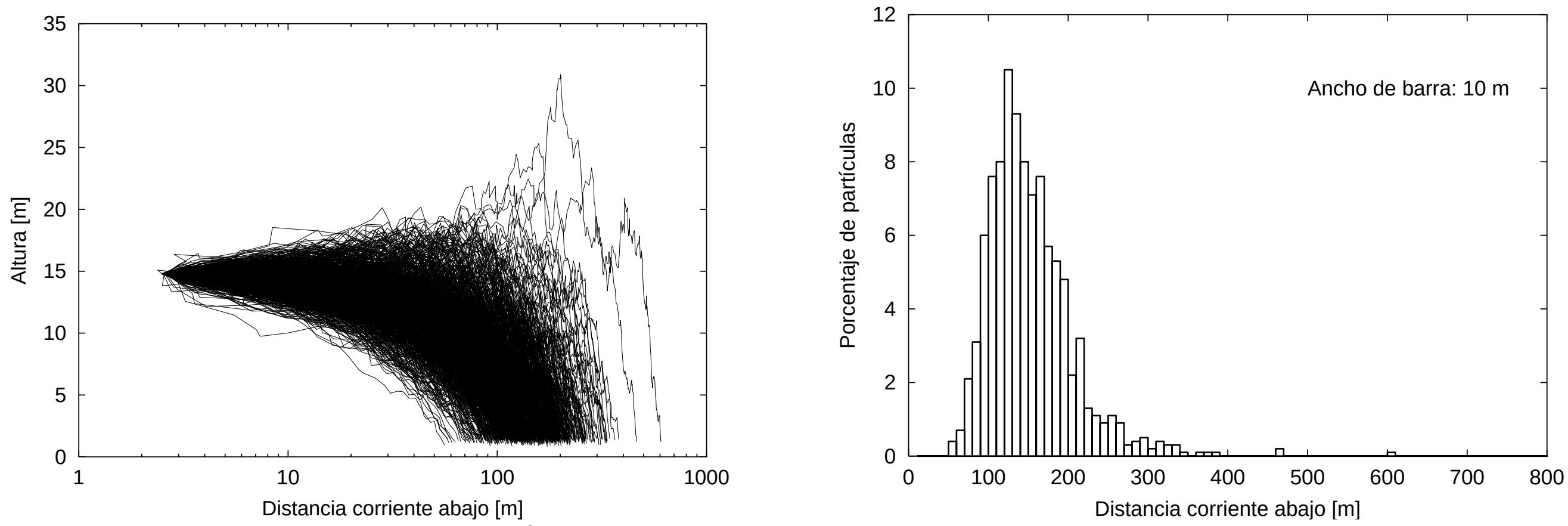

Fig. 1: (a) Trayectorias de partículas liberadas desde 15m. (b) Histograma correspondiente de deposición. El pico de la distribución está en la clase centrada en 125 m (ancho de clase 10 m); se encuentran partículas hasta una distancia de 605.4 m corriente abajo según la dirección del viento; a 200 m ha caído el 86.4 % del material particulado simulado.

diferencias con todos los parámetros estadísticos evaluados dentro de un 12 %.

En este trabajo el rango de interés se centra en partículas de 50, 75 y 100 µm emitidas por chimeneas de 50 m de altura. Básicamente se analizaron tres casos: 1) *velocidad inicial*: velocidad de escape a través de la chimenea de 10 m $s^{-1}$; 2) *velocidad inicial nula*: partícula liberada a la salida de la chimenea sin velocidad apreciable de emisión; 3) *sin flotación*: caso 1) despreciando la densidad del aire. En todos los casos la velocidad del viento es la misma. Los casos 1) y 2) tienden a evaluar el efecto de la velocidad de eyección, cantidad de difícil estimación práctica. El caso 3) evalúa el efecto de la flotabilidad de las partículas a simular. La motivación de este caso se origina en que los experimentos de Vesovic et al. (2001), aunque no lo justifican explícitamente, no consideran la flotabilidad de las partículas en el aire. Para el presente trabajo resultó de interés evaluar su efecto en la dinámica (fuertemente caótica) de estas partículas, debido a que la altura de lanzamiento es significativamente mayor (las partículas tienen mucho más tiempo de permanencia en el aire).

Los resultados se muestran en la Fig, 2, donde se grafican los porcentajes de partículas caídas, agrupados cada 50 m, para los tres casos de los tres tipos de PMc analizado (densidad de las partículas 5.5 g $cm^{-3}$). Las curvas de los casos 1) y 2) muestran gran coincidencia, a pesar de tener velocidades iniciales completamente diferentes (para 100 µm las curvas son indistinguibles); en efecto un test chi-cuadrado realizado sobre las distribuciones indica coincidencia entre ambas con gran nivel de significación. Como es de esperar para la densidad indicada, despreciar la flotabilidad no produjo resultados significativos. Algunos parámetros estadísticos del caso 1) (moda, mediana y distancia para la cual se ha depositado el 80% del PMc) se muestran en la Tabla 1.

Estos resultados (y otros obtenidos con diferentes velocidades iniciales) sugieren que, dentro de los parámetros estudiados, la velocidad del viento maneja la dispersión de las partículas. Para mostrar el efecto del cambio en la velocidad del viento, la Fig. 2, muestra también la curva de deposición del caso 1) con la intensidad del viento aumentada hasta un valor en donde comienzan a apreciarse cambios significativos. Para el PMc de 50 µm el porcentaje resulta ser aproximadamente del 30 %, para 75 µm del 25 % y para 100 µm del 20 %.

La densidad del PMc es sumamente heterogénea, por lo tanto se realizaron experimentos con partículas de diferentes densidades, obteniéndose el mismo tipo de resultado. Sin embargo, para densidades de 2 g $cm^{-3}$ (similares a las usadas por Vesovic, et al) despreciar la densidad del aire en partículas de 50 μm comienza a evidenciar sobreestimación en los máximos de concentración (5.7 %) y subestimación de la moda (17%), indicando que para este tipo de simulaciones la flotabilidad no es una fuerza despreciable.

Además, se han efectuado simulaciones variando la forma de las partículas. Por ejemplo, para el caso 1) (el más realista), se varió aleatoriamente a lo largo de las simulaciones el diámetro de las partículas en ± 50%, para considerar posibles irregularidades en sus formas esferoidales. El resultado no revela diferencias significativas con el caso 1). Otras simulaciones con tamaño fijo y factores de forma diferentes a la esfera producen variaciones importantes en la deposición, en acuerdo con Vesovic et al. (2001), produciendo mayor dispersión en la deposición y alejamiento del pico de

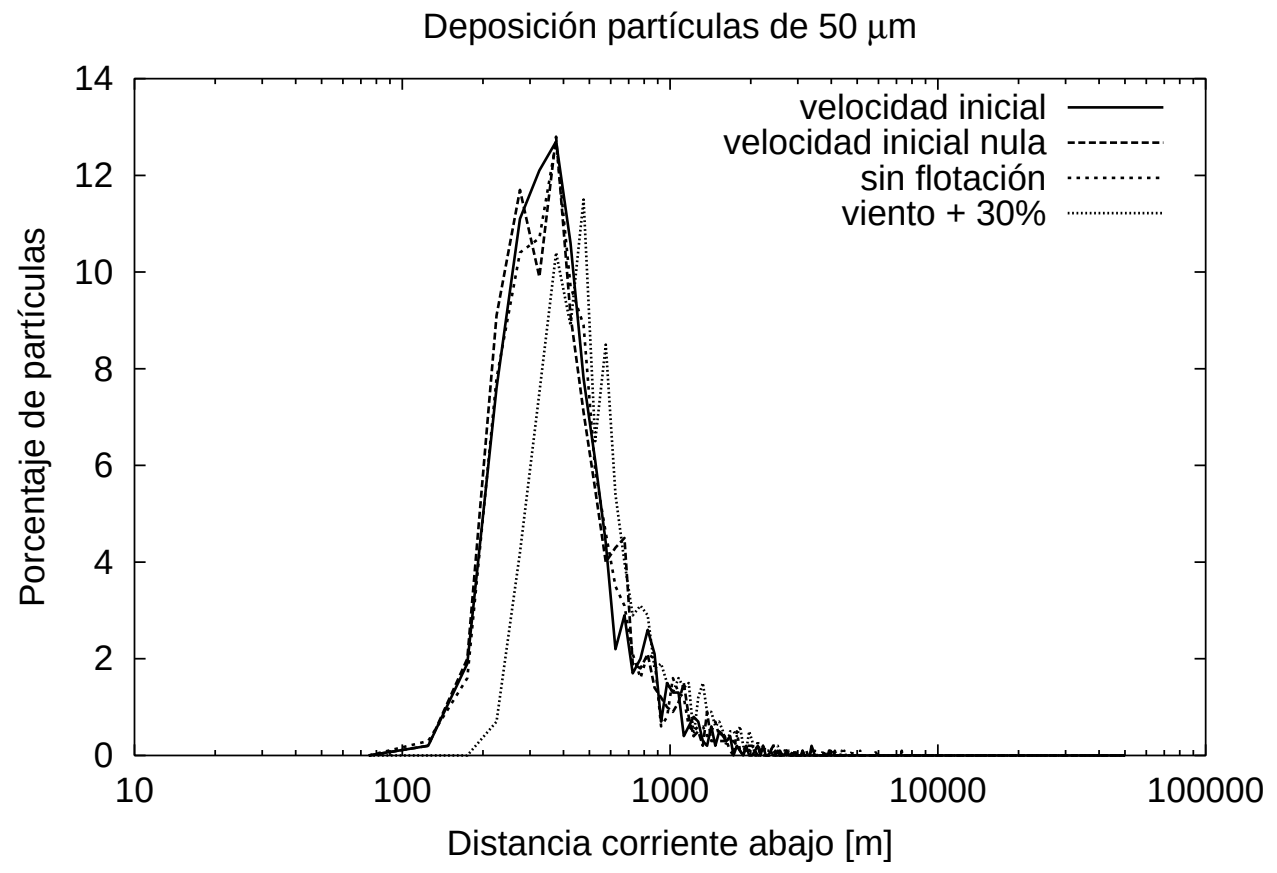

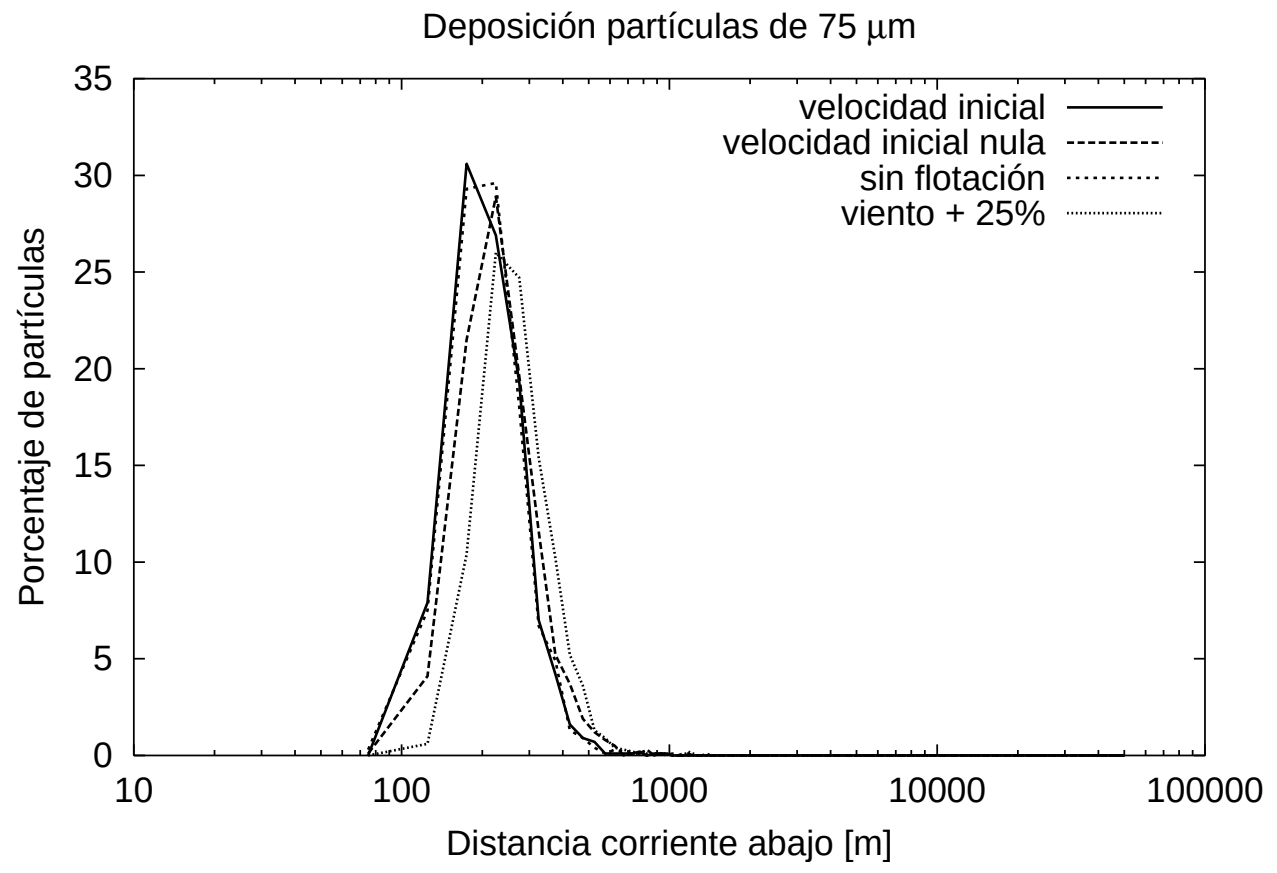

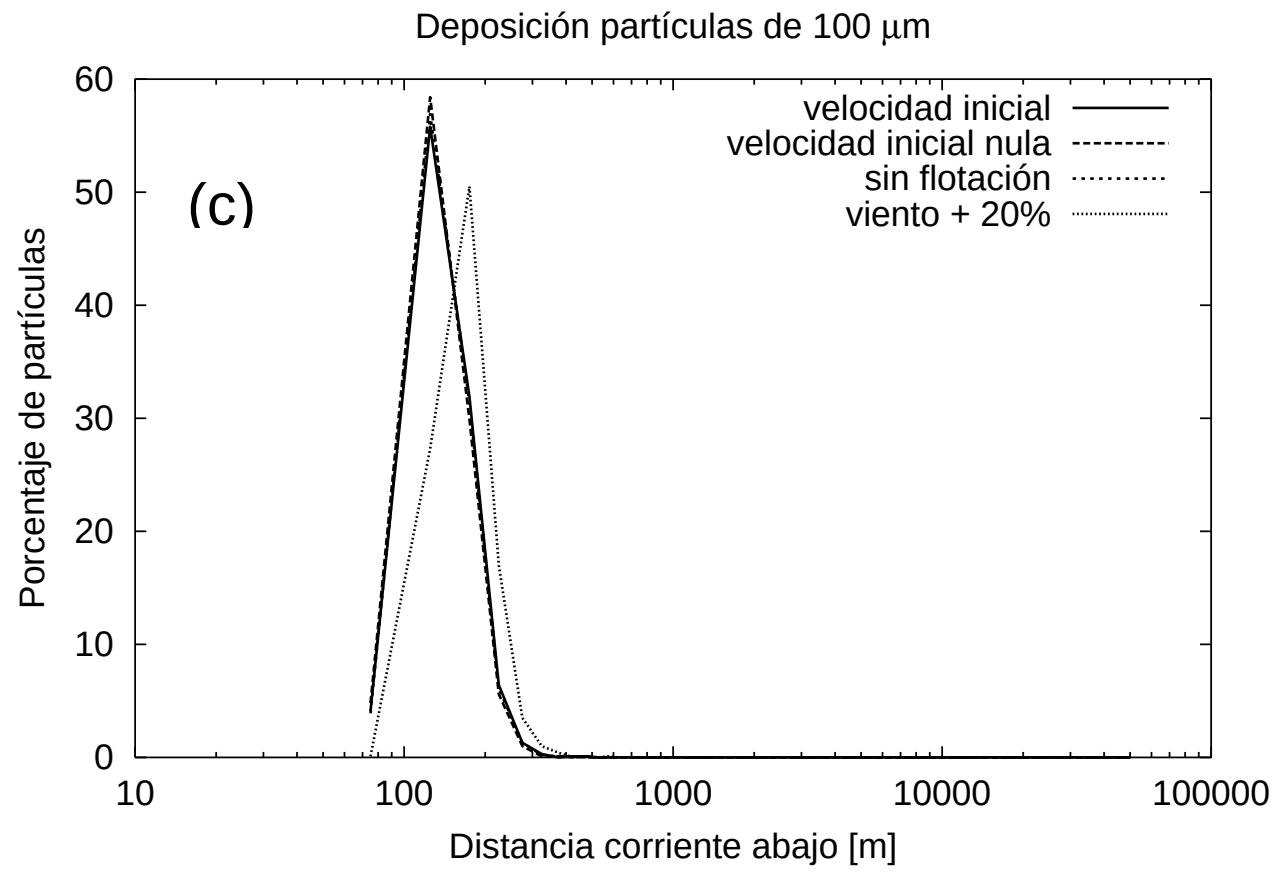

Tabla1. Estadística relevante de las distribuciones anteriores (Fig. 2 a, b y c), para el caso 1); unidades en metro.

| | Moda | Mediana | X<80 % |
|---|---|---|---|
| 50 μm | 445.64 | 523.44 | 1050.00 |
| 75 μm | 242.79 | 245.98 | 250.00 |
| 100 μm | 170.10 | 170.40 | 210.00 |

Fig. 2: Curvas de deposición (tres casos) para partículas de 50 μm (a), 75 μm (b) y 100 μm (c). Además se muestra la deposición aumentando la intensidad del viento para el caso 1) en los porcentajes indicados.

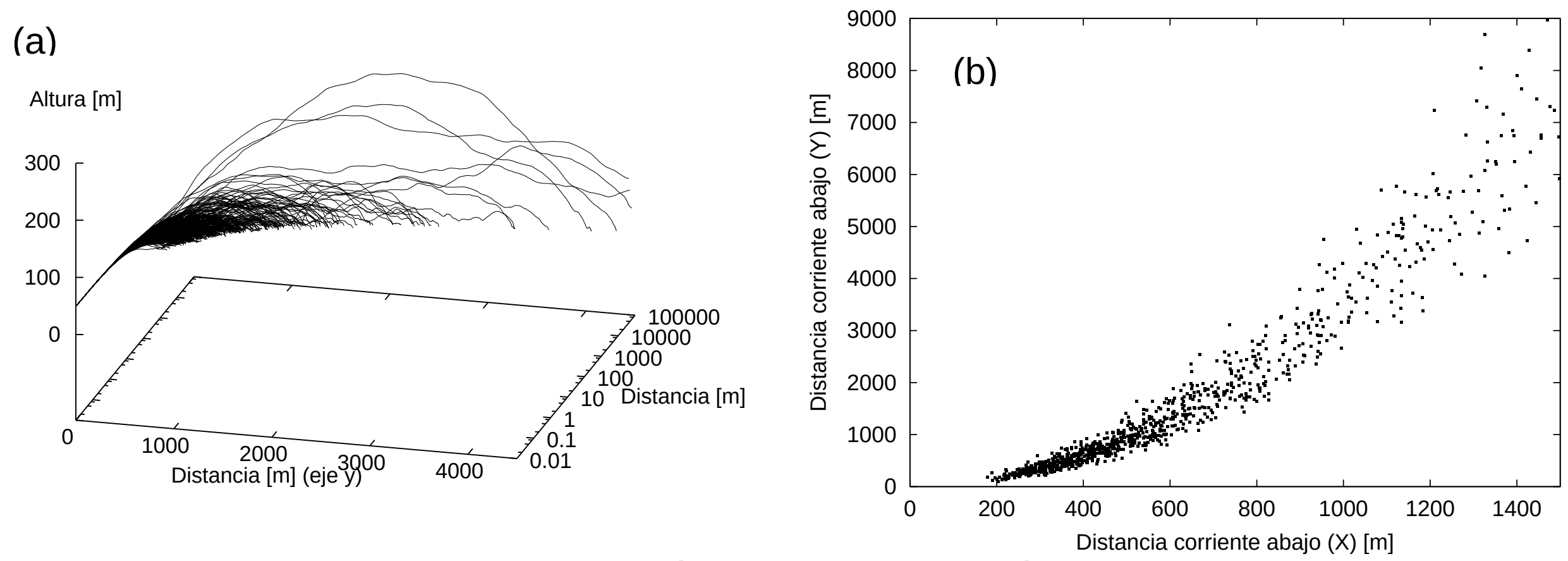

Fig. 3: Panel a), trayectorias de algunas partículas de la simulación 3D, con viento rotante. Panel b) plano de deposición de las mismas partículas.

concentración; este efecto es más apreciable en las partículas de 100 μm. Con objeto de recrear trayectorias tridimensionales, se presentan simulaciones del caso 1), utilizando mil partículas de 50 μm, donde el viento presenta una rotación hacia el eje *Y*; se considera también turbulencia sobre este eje. La Fig. 3 a), muestra las trayectorias de algunas partículas; la Fig. 3 b) presenta el plano de deposición correspondiente (coordenadas *x*, *y* de las partículas en el instante del alcanzar la altura mínima de integración).

**CONCLUSIONES**

Este trabajo reporta los primeros resultados de simulaciones de PMc emitido por chimeneas industriales realizadas con un código propio. El integrador Bulirsch y Stoer resulta ser sumamente preciso y rápido también para este tipo de cálculos, a pesar de las dificultades inherentes a la turbulencia. Las simulaciones anteriores demandan casi 1 minuto de CPU (densidades 5 g $cm^{-3}$) a 25 - 50 minutos de CPU (densidades 2 g $cm^{-3}$), en una notebook Pentium III de 850Mhz. La subdivisión automática del paso de integración, hasta alcanzar la tolerancia deseada (usualmente ~ $10^{-6}$), evita problemas de convergencia relacionados con la longitud de $h_1$ (una de las principales ventajas del método), el cual queda físicamente limitado por el tiempo lagrangiano mínimo para la turbulencia. Simulaciones de partículas entre 50 - 100 μm indican que, a los efectos de la deposición, las velocidades de emisión del PMc no son tan importantes frente a la velocidad del viento, parámetro que maneja la deposición de las partículas. En el rango de PMc estudiado, despreciar la flotabilidad en partículas de 5 g $cm^{-3}$ no introduce efectos apreciables, mientras que para partículas de 50 μm y 2 g $cm^{-3}$ ya comienzan a notarse diferencias significativas en las curvas de deposición. Para partículas esféricas, los resultados sugieren que la variación irregular del área aerodinámica no es un factor de importancia, presentando curvas de deposición similares a los de una esfera de radio fijo. Sin embargo, apartamientos de la esfera, producen marcadas variaciones en la concentración del PMc depositado. Se espera de aquí en más continuar explorando el espacio de parámetros del problema y diversificar los casos de aplicación.

**AGRADECIMIENTOS**

Los autores agradecen al Dr. Adrián Brunini por permitir el uso de recursos informáticos del GCP-UNLP para realizar simulaciones de prueba.

**REFERENCIAS**